\documentclass[aps,pra,reprint,nofootinbib,twocolumn,superscriptaddress,showpacs,showkeys,longbibliography]{revtex4-1}
\usepackage{eurosym}
\usepackage{amsmath,amssymb,amstext}
\usepackage[usenames,dvipsnames]{color}
\usepackage{graphicx}
\usepackage{braket}
\usepackage{natbib}
\usepackage{comment}
\usepackage{dcolumn}
\usepackage[english]{babel}
\usepackage{wasysym}
\usepackage[colorlinks,bookmarks=false,citecolor=blue,linkcolor=red,urlcolor=blue]{hyperref}
\newcommand{\mathsym}[1]{{}}
\newcommand{\unicode}[1]{{}}

\begin{document}
\title{Non-equilibrium quantum phase transition in a spinor quantum gas in a lattice coupled to a membrane}

\author{Xingran Xu}
\affiliation{Department of Physics, Zhejiang Normal University, Jinhua, 321004, China}
\affiliation{Shenyang National Laboratory for Materials Science, Institute of Metal Research, Chinese Academy of Sciences, Shenyang, China}
\affiliation{ School of Materials Science and Engineering, University of Science and Technology of China, Hefei, China}
\author{Zhidong Zhang}
\affiliation{Shenyang National Laboratory for Materials Science, Institute of Metal Research, Chinese Academy of Sciences, Shenyang, China}
\affiliation{ School of Materials Science and Engineering, University of Science and Technology of China, Hefei, China}
\author{Zhaoxin Liang}
\email{The corresponding author: zhxliang@gmail.com}
\affiliation{Department of Physics, Zhejiang Normal University, Jinhua, 321004, China}

\date{\today}

%Abstract

\begin{abstract}
Recently, a novel kind of hybrid atom-optomechanical  system, consisting of atoms in a lattice coupled to a membrane, has been experimentally realized [Vochezer {\it et al.,} Phys. Rev. Lett. \textbf{120}, 073602 (2018)], which promises a viable contender in the competitive field of simulating non-equilibrium many-body physics. Here we are motivated to investigate a spinor Bose gas coupled to a vibrational mode of a nano-membrane, focusing on analyzing the role of the spinor degrees of freedom therein. Through an adiabatic elimination of the degrees of freedom of the quantum oscillator, we derive an effective Hamiltonian which reveals a competition between the force localizing the atoms and the membrane displacement. We analyze the dynamical stability of the steady state using Bogoliubov-de Gennes approach and derive the stationary phase diagram in the parameter
space. We investigate the non-equilibrium quantum phase transition from a localized symmetric state of the atom cloud to a shifted symmetry-broken state, where we present a detailed analysis of the effects of the spin degree of freedom. Our work presents a simple way to study the effects of the spinor degree of freedom on the non-equilibrium nonlinear phenomena that is complementary to ongoing experiments on the hybrid atom-optomechanical system.
\end{abstract}

\maketitle

\section{Introduction}
In recent years, the hybrid atom-optomechanical systems~\cite{Vochezer2018,Christoph2018,Ritsch2018,Mann2018}, where a membrane is coupled to ultra-cold quantum gases, have attracted considerable interests as a novel and versatile alternative to more conventional optomechanical setups. Combining mechanical oscillators and ultra-cold atoms, such hybrid systems~\cite{Camerer2011,Vogell2013,Bennett2014,Vogell2015,Jockel2015,Moller2017} provide opportunities for cooling, detection and quantum control of mechanical motion, with applications in precision sensing, quantum-level signal transduction, as well as for fundamental tests of quantum mechanics~\cite{Marquardt2009,Aspelmeyer2014,Sudhir2017,Harris2017,Vaidya2018}. For example, state-of-the-art optomechanics is nowadays able to
realize optical feedback cooling of the mechanical oscillator to its quantum-mechanical ground state~\cite{Christoph2018}. Being intrinsically non-equilibrium, such hybrid mechanical atomic system further provides a natural setting for non-equilibrium many-body quantum systems. Adding phononic degrees of freedom to the optical lattice toolbox~\cite{Jaksch2005,Bakhtiari2015},  it also opens new routes to mimic the lattice vibrations and quantum simulations of phonon dynamics in realistic solid materials~\cite{Gao2019}.

Building on above development, further accounts of the spinor degree of freedom of the atom part, which is a key ingredient playing
out in modern physics, are expected to reveal exceptionally rich physics in hybrid atom-optomechanical  systems. In this work, we are motivated to
study a spinor hybrid atom-optomechanical setup that consists of a membrane coupled to spinor ultracold quantum gases. There, the light-mediated coupling between the atoms and the membrane is non-resonant, allowing for adiabatic elimination of the degree of freedom of the quantum oscillator. The resulting Hamiltonian can be regarded as a nonlinear quantum system in periodic potentials. Solving the Bogoliubov-de Gennes equations, we derive the dynamical stability phase diagram for this system in the parameter space. As the atom-membrane coupling is tuned via controlling the
laser intensity, a non-equilibrium quantum phase transition
(NQPT) is induced between a localized symmetric state and a symmetry-broken quantum many-body state exhibiting a shifted
cloud-membrane configuration. Finally, we discuss how the stationary-state phase can be probed through the elementary excitations of the model system. We believe our model provides a simple way to study the non-equilibrium nonlinear phenomena that is complementary to ongoing experiments on the hybrid atom-optomechanical systems.

The emphasis and value of the present work are to provide a theoretical model, i.e. an extended two-component Gross-Pitaevskii equation coupled to a quantum harmonic oscillator in describing the hybrid mechanical-atomic system, which at the mean-field level captures the key physics regarding the interplay of quantum many-body physics, non-equilibrium nature and the spinor degree of freedom. Our study builds on recent progress in engineering the optomechanical  coupling $\lambda$ in experiments~\cite{Vochezer2018,Mann2018}. For vanishing intrinsic optomechanical coupling $\lambda\rightarrow 0$, our model reduces to the equilibrium two-component condensates which have been intensively explored both theoretically and experimentally in the context of ultracold quantum gases~\cite{SpinRev1,Mivehvar2019,Ostermann_2019}. Note that our previous work~\cite{Gao2019} has obtained the steady-state phase diagram for the \textit{one-component}  hybrid mechanical-atomic system, which has extended studies on the steady-state phases from the superfluid regime into the full parameter regimes. In this work, we further account for the spinor degree of freedom of the atom part. 
The three work together will provide a complete description of  the steady-state phase diagram of the hybrid mechanical-atomic system experimentally motivated by Ref.~\cite{Jockel2015,Vochezer2018,Mann2018}. We hope the theoretical model proposed in this work can serve as an alternative model to study the spinor non-equilibrium nonlinear phenomena in a highly controllable way.

The paper is organized as follows. In Sec.~\ref{sec:model}, we briefly describe the model system and corresponding mean-field treatment.
In Sec.~\ref{sec:stability}, we revisit the dynamical stability analysis of
the stationary state by means of Bogoliubov-de Gennes
approach and derive the dynamical stability phase diagram of the model system in the parameter
space. Sec.~\ref{sec:NQF} presents detailed analysis on the non-equilibrium quantum phase transition, in particular, the role of the spinor nature of the atomic gas on the quantum phase. We conclude in Sec.~\ref{sec:conc}. 

\section{Model Hamiltonian}\label{sec:model}
In this work, we consider a spinor hybrid mechanical-atomic system consisting of a membrane in a single-sided optical cavity, i.e. one mirror of the cavity is designed to reflect incident light on resonance and forms a standing wave in front of the cavity, in which a spinor Bose-Einstein condensate (BEC) can be trapped. Our setup is of immediate relevance in the context of experiments for the one-component hybrid mechanical-atomic system \cite{Vochezer2018,Christoph2018,Ritsch2018,Mann2018}. Furthermore, the spin degree of freedom can be encoded by two atomic internal states or sub-lattices \cite{SpinRev1}. Our goal is to find a nonequilibrium quantum phase transition from a localized symmetric state of the atom to a shifted symmetry broken, in particular, focus on the spin degree of freedom's effects on the phase transition.

The atom part of our model consists of a two-component BEC in an optical lattice along the $x$-direction, wheres the model system is uniform in the other two directions. To be specific, we consider $^{87}$Rb and choose the internal states of $\left|F=1,m=0\right\rangle$ and $\left|F=1,m=-1\right\rangle$ as a pseudo-spin-$1/2$ system. As such, the freedom along the $y$- and $z$- directions decouples from the $x$-direction, leading to the realization of a quasi-one-dimensional geometry. Within the mean-field approximation, the order parameter for the condensate can be described by a two-component time-dependent wave function $\Psi=[\psi_1,\psi_2]^T$, which dynamics can be well described by the two-component Gross-Pitaevskii  (GP) equations, i.e.,
\begin{eqnarray}
i\hbar\frac{\partial}{\partial t}\psi_{1}&=&-\hbar\omega_{R}\partial_{x}^{2}\psi_{1}+V\sin^{2}(x)\psi_{1}+\hbar\Omega\psi_{2}+gN\left|\psi_{1}\right|^{2}\psi_{1}\nonumber\\
&+&g_{12}N\left|\psi_{2}\right|^{2}\psi_{1}-2\sqrt{N}\lambda\alpha_1\sin(2x)\psi_{1}\label{GP1}\\
i\hbar\frac{\partial}{\partial t}\psi_{2}&=&-\hbar\omega_{R}\partial_{x}^{2}\psi_{2}+V\sin^{2}(x)\psi_{2}+\hbar\Omega\psi_{1}+gN\left|\psi_{2}\right|^{2}\psi_{2}\nonumber\\
&+&g_{12}N\left|\psi_{1}\right|^{2}\psi_{2},\label{GP2}
\end{eqnarray}
with $V$ being lattice potential strength, $N$ the number of the condensed atoms,  $\hbar\omega_R$ is the kinetic energy,  $\Omega$ denotes Rabi frequency,  $g$, and $g_{12}$ label inter-atomic and intra-atomic interactions  respectively. Here, the coupling between the atoms and the membrane labeled by $\lambda$ can be obtained with a Born-Markov approximation by adiabatically eliminating the light field~\cite{Mann2018,Gao2019}. The $\alpha_1$ is referred  to the real part of the complex amplitude $\alpha$ of a coherent state (see Eq. (\ref{CoherentE})). Note that going beyond the GP equations (\ref{GP1}) and (\ref{GP2}) to fully include the quantum and thermal fluctuations of the quantum field is beyond the scope of this work.

The motion of the membrane can be treated as a one-dimensional quantum oscillator with frequency $\Omega$, $H_{\text{m}}=\hbar\Omega_m a^\dagger a$. Within the mean-field framework, we are interested in the dynamics of the mean value of the field operator $a$ under the coherent ansatz $\langle a\rangle= \alpha$. The equation of motion of $\alpha$ can be written as 
\begin{equation}
i\frac{\partial}{\partial t}\alpha=(\Omega_{m}-i\gamma)\alpha-\sqrt{N}\lambda\int dx\sin(2x)\left|\psi_{1}\right|^{2}.\label{CoherentE}
\end{equation}
Here the $\gamma$ represents a phenomenological damping rate and the membrane is coupled to one component of the two-component BEC. Note that $\alpha=\alpha_1+i\alpha_2$ in Eq.~(\ref{CoherentE}) is a complex number ($\alpha_1$ and $\alpha_2$ being its real and imaginary part respectively) and plays a role of the order parameter for the membrane. The physical meaning of $\alpha$ can be regarded as the displacement of the membrane around its equilibrium. In more details, $\alpha=0$ denotes an incoherent vibration state of  the membrane, wheres $\alpha\neq 0$ denotes a coherent vibration. 

The stationary-state phase diagram of the spinor hybrid atom-optomenchanical  system described by Eqs.~(\ref{GP1}) and (\ref{GP2}) is determined by five parameters: the lattice strength $V$, the coupling constant $\lambda$ between the atom and membrane, the inter- and intra-atomic interactions $g$ and $g_{12}$ and the Rabi frequency $\Omega$. Note that there is a quantum phase transition for $\lambda=0$ in the context of equilibrium ultra-cold atomic BEC~\cite{Abad2013} and dissipative polariton BEC \cite{Xu2017}: $g_{12}>g+2\Omega/n$ the system turns from unpolarized phase to polarized phase for order parameter $\langle\sigma_z\rangle=n_1-n_2$ is zero or nonzero. 
In what follows, we address how the non-equilibrium nature of the model system, i.e. $\lambda\neq 0$, can affect the above quantum phase transition.

To motivate our discussion of effects of the spinor degree of freedom on the phase transition,  one notices two important features with respect to the framework of Ref. \cite{Mann2018}: first, the spinor
degree of freedom of our model system is encoded in the two-component order parameters $\left[\psi_1,\psi_2\right]$; second, the membrane is coupled to the superposition of both the density and spin-density of the BEC, which will inevitably couple to excitations in the density and spin-density fluctuations. This further justifies our motivation of focusing on effects of the spinor degree of freedom on the phase transition.

From Eqs. (\ref{GP1}) and (\ref{GP2}), the key physical picture behind the non-equilibrium quantum phase transition can immediately be stated as follows: there exist two different kinds of periodic potentials, which dynamically compete with each other, depending on the back action of the membrane on the atoms, and thus on the collective behavior of the atoms. We are interested in the tight-binding limit, where the lattice is so strong that the BEC system can be considered as a chain of trapped BECs that are weakly linked.

\section{Stability of the hybrid mechanical-atomic system}\label{sec:stability}

The main goal of this work is to investigate the non-equilibrium quantum phase transition in a spinor quantum gas in a lattice coupled to a membrane. Before proceeding, we remark that the stationary states of a periodically-trapped quantum gas is represented by a Bloch wave \cite{Wu2001,LiWu2015}, i. e. a plane wave with periodic modulation of the amplitude. 
One unique feature in the system of the quantum gas in optical lattices coupled to a membrane is dynamical instability \cite{Wu2001,Wu2003}, which does
not exist in the absence of either atomic interaction. In more detail, some of the Bloch waves can be dynamically unstable against certain perturbation modes $q$ only when both factors are present. By dynamical instability, we mean that small deviations from a state grow exponentially in the course of time evolution. Therefore, as a first step, it is important to check whether the Bloch wave itself is stable against
weak perturbations, which is the aim of this section.

\begin{figure}
  \centering
  % Requires \usepackage{graphicx}
  \includegraphics[width=0.5\textwidth]{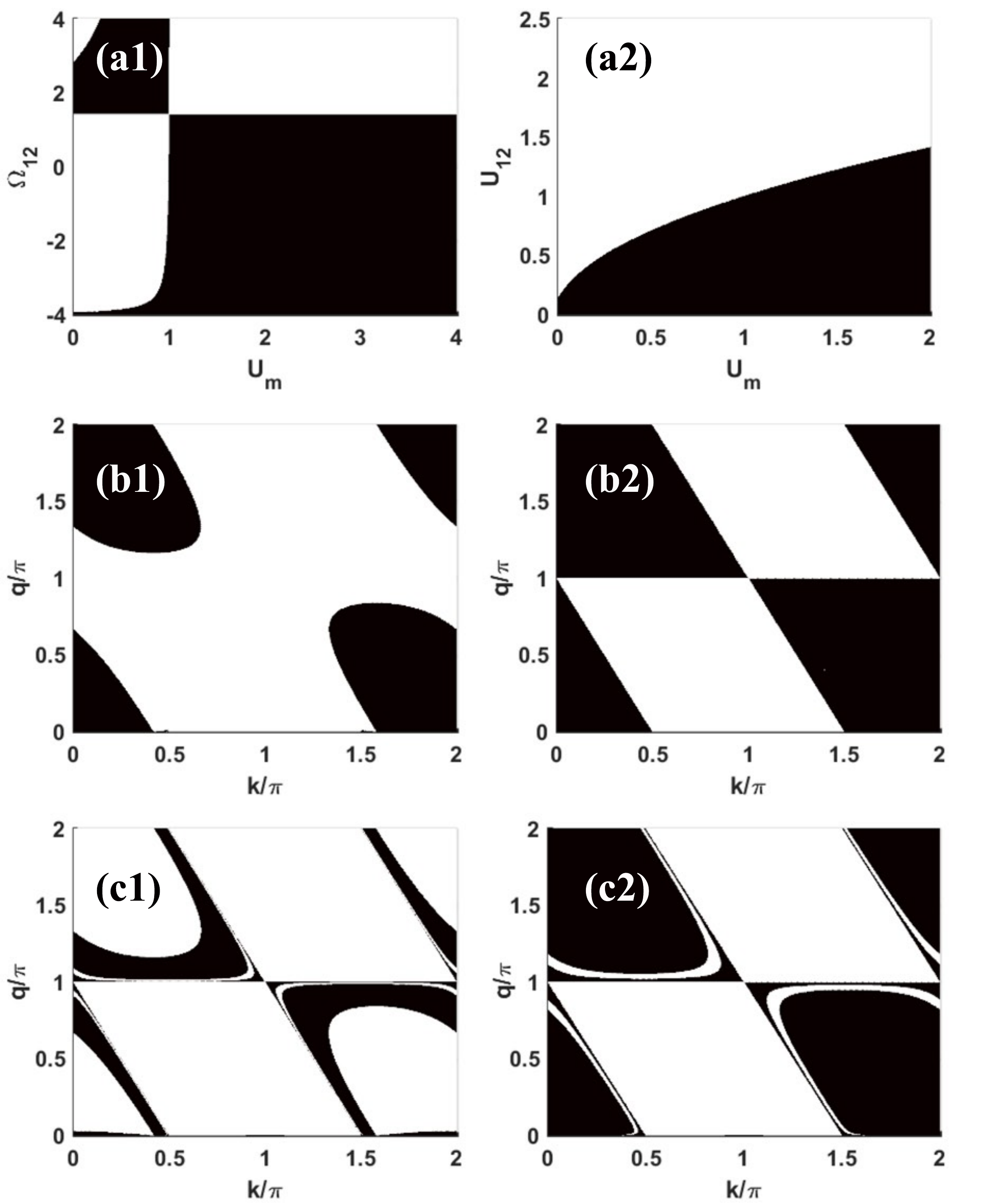}\\
  \caption{Dynamical instability of a spinor quantum gas in a lattice coupled to a membrane. In the black-color regions, imaginary
parts of dispersion spectrum for excitations of a Bloch wave are zero or negative, representing dynamical stability regions; while in the white-color regions, imaginary
parts of dispersion spectrum for excitations of a Bloch wave are positive, suggesting dynamical instability of the condensate. Parameters are chosen as: $K_{1}$=$K_{2}$=1, and  (a1): $k$=$\pi$/4, $q$=0, $U_{12}/U$=1; (a2): $k$=$\pi$/4, $q$=0, $\Omega_{12}/Un$=0; (b1):$U_{12}/U$=1, $ \Omega/Un$=0.5, $U_m/U$=1.2; (b2): $U_{12}/U$=1, $ \Omega/Un$=0.5, $U_m/U$=1.0; (c1):  $U_{12}/U$=1, $ \Omega/Un$=0.5, $U_m/U$=0.8; and (c2): $U_{12}/U$=1, $ \Omega/Un$=2, $U_m/U$=0.8.}\label{MI}
\end{figure}

We are interested in the parameter regime of strong lattice strength within the framework of the tight-binding approximation. Directly following Ref. \cite{Liang2008,Smerzi2001,Jaksch1998}, we proceed to expand the order parameters of $[\psi_{1},\psi_{2}]^T$ in the Wannier basis and keep only the lowest vibrational states as follows
\begin{eqnarray}
\psi_{1}&=&\sqrt{N}\sum_{m}a_{m}\left(t\right)\phi\left(x-x_{1,m}\right),\label{TB1}\\
\psi_{2}&=&\sqrt{N}\sum_{m}b_{m}\left(t\right)\phi\left(x-x_{2,m}\right),\label{TB2}
\end{eqnarray}
with  $\phi\left(x-x_{i,m}\right)$ is a Wannier function at the $m$ sites and $x_{i,m}$ represents the central position of $i$ component at $m$ site. 

In the similar way,  Equation (\ref{CoherentE}) can be rewritten as under the tight-binding approximation,
\begin{equation}
i\frac{\partial}{\partial t}\alpha=(\Omega_{m}-i\gamma)\alpha-Q\sum_{n}\left|a_{m}\right|^{2}, \label{MIQ}
\end{equation}
with
\begin{equation}
Q=\lambda N^{3/2} \int dx\sin(2x)\left|\phi\left(x-x_{1,m}\right)\right|^2.
\end{equation}
We focus on the stationary of the membrane by letting $\partial \alpha/\partial t=0$ in Eq. (\ref{MIQ}). In such, we can obtain the value of the steady state $\alpha_0$ and then
 the coupling strength between BEC and the membrane is connected to the real part of $\alpha_0$. By substituting the steady state of Eq.~(\ref{MIQ}) into  Eq.~(\ref{GP1}), we arrive at
  \begin{eqnarray}
  i\hbar\frac{\partial \psi_{1}}{\partial t}&=&-\omega_{R}\partial_{x}^{2}\psi_{1}+V\sin^{2}(x)\psi_{1}+gN\left|\psi_{1}\right|^{2}\psi_{1}+\Omega\psi_{2}\nonumber\\
  &+&g_{12}N\left|\psi_{2}\right|^{2}\psi_{1}-\frac{2\sqrt{N}\lambda Q}{\Omega_m} \sum_{m}\left|a_{m}\right|^{2}\sin(2x)\psi_{1}.\label{EffGP1}
  \end{eqnarray}
Two properties of the effects of the back action of the membrane on the quantum gas can immediately be stated based on Eq. (\ref{EffGP1}): (i)
this effective lattice with the renormalized lattice strength shares the same periodicity;
(ii) its lattice site location is shifted from that of the original lattice, \(x_m^{(0)}=m a_\text{L}\) (\(m=0,1,2\text{...}\)), to \(x_m=m a_\text{L}+\delta\) by \(\delta\). 
The back action of the membrane on the quantum gas is to provide the competition between  the optical lattice, trying to localize the atoms at the minima, and the membrane displacement which tries to shift the atoms.

Furthermore,  by plugging Eqs. (\ref{TB1}) and (\ref{TB2}) into Eqs. (\ref{EffGP1}) and (\ref{GP2}), we can obtain the discrete nonlinear Schr\"odinger equations as follows
\begin{eqnarray}
i\hbar\frac{\partial}{\partial t}a_{m}&=&-K_{1}\left(a_{m-1}+a_{m+1}\right)+\Omega_{12}b_{m}\nonumber \\
&+&\left(\epsilon_{1,m}+U_{1,m}\left|a_{m}\right|^{2}+U_{12}\left|b_{m}\right|^{2}\right)a_{m},\label{DS1}\\
i\hbar\frac{\partial}{\partial t}b_{m}&=&-K_{2}\left(b_{m-1}+b_{m+1}\right)+\Omega_{12}b_{m} \nonumber \\
&+&\left(\epsilon_{2,m}+U_{2,m}\left|b_{m}\right|^{2}+U_{12,m}\left|a_{m}\right|^{2}\right)b_{m}.\label{DS2}
\end{eqnarray}
with
\begin{equation} 
K_i=-\int \left[\omega_R\frac{\partial}{\partial x}\phi_{i,m}\frac{\partial}{\partial x}\phi_{i,m+1}+V\sin^2 x  \phi_{i,m}\phi_{i,m+1}\right]dx
\end{equation}
 is the nearest neighbor hopping,
\begin{equation}
\epsilon_{i,m}=\frac{1}{2}\int \left[\omega_R\left|\frac{\partial}{\partial x}\phi_{i,m}\right|^2+V\sin^2 x \left| \phi_{i,m}\right|^2\right]dx
\end{equation}
 is the effective potential on every site, $U_{2,m}=\frac{gN}{2}\int\left|\phi_{2,m}\right|^{4}dx$
is the on-site atomic collisions, $U_{12,m}=gN\int\left|\phi_{1,m}\right|^{2}\left|\phi_{2,m}\right|^{2}dx$  and $U_{1,m}$ can be adjusted by $\alpha_0$ and $\lambda$ with
\begin{eqnarray}
U_{1,m}&=&\frac{gN}{2}\int\left|\phi_{1,m}\right|^{4}dx \nonumber\\
&-&\frac{2\sqrt{N}\lambda Q}{\Omega_m} \int \sum_{n}\sin(2x) \left| \phi_{1,m}\right |^2 dx.
\end{eqnarray}

The condition of the dynamical  instabilities of  Bloch waves solution can be determined based on Eqs. (\ref{DS1}) and (\ref{DS2}) as follows: we start from the standard decomposition of the  wave functions into the steady-state solution labelled by the Bloch wave number $k$ and a small fluctuating term with $q$ being also a kind of Bloch wave number
\begin{eqnarray}
a_{n}&=&\left(\psi_{10}+u_{1}e^{iqm-i\omega t}+v_{1}^{*}e^{-iqm+i\omega t}\right)e^{ikm-i\mu t},\label{BP1}\\
b_{n}&=&\left(\psi_{20}+u_{2}e^{iqm-i\omega t}+v_{2}^{*}e^{-iqm+i\omega t}\right)e^{ikm-i\mu t},\label{BP2}
\end{eqnarray}
with $\psi_{10}=\psi_{20}=\sqrt{n_0}/2$ and considering one site with $U_{12,m}=U_{12}$, $U_{2,m}=U$ and $U_{1,m}=U_m$. Substituting Eqs. (\ref{BP1}) and (\ref{BP2}) into Eqs. (\ref{DS1}) and (\ref{DS2}) and retaining only first-order terms of fluctuation, we obtain at each momentum $k$ the Bogoliubov-de Gennes (BdG) equation $M_{k}U_k=\hbar\omega_k U_k$ with $U_k=(u_1,v_1,u_2,v_2)^T$. Here the $M_k$ in the matrix form reads as
\begin{eqnarray}
M=\left(\begin{array}{cccc}
h_{1} & U_{m}n & U_{12}n+\Omega_{12} & U_{12}n\\
-U_{m}n & -h_{1} & -U_{12}n & -U_{12}n-\Omega_{12}\\
U_{12}n+\Omega_{12} & U_{12}n & h_{2} & Un\\
-U_{12}n & -U_{12}n-\Omega_{12} & -Un & -h_{2}
\end{array}\right).\label{MIMatrix}
\end{eqnarray}
Here, $h_a$ and $h_b$ are diagonal terms of Matrix, reading
\begin{equation}
h_{i}=2K_{i}\left[\cos(k+q)+\cos(k)\right]+nU_{i}-\Omega_{12}.
\end{equation}
In some parameter regions,  the imaginary parts of eigenvalues of Eq. (\ref{MIMatrix}) are positive and the condensate wave functions with the form of Bolch waves
become to be dynamical instability, i.e.  the density modulations grow in time exponentially. Stability phase diagrams of the spinor quantum gas in a lattice coupled to a membrane in the tight-binding limit are plotted in Fig.~\ref{MI}. In the white-color regions of Fig. \ref{MI}, the imaginary
parts of dispersion spectrum for excitations of a Bloch wave are positive, suggesting dynamical instability of the condensate.
These regimes correspond to effectively attractive nonlinearity of two-component GP equations as explained in Refs. \cite{Xu2017,Smirnov2014}. Consequently, the growth
of the spatial density modulations is supposed to lead to the formation
of steady states with modulated density, which goes beyond the scope of current work. In what follows, we restrict our consideration to the
dynamics of nonlinear waves propagating on a dynamically
stable condensate background. Therefore we make
sure that the parameters of the system always satisfy the dynamical stability condition.

\section{Non-equilibrium quantum phase transition}\label{sec:NQF}

The goal of this section is to investigate the non-equilibrium quantum phase transition based on Eqs. (\ref{GP1})-(\ref{CoherentE}). At the heart of our solution of non-equilibrium dynamics of the spinor hybrid mechanical-atomic system is that  (i) an elimination of the degrees of freedom the membrane, leading to an effective Lagrangian where the parameters are significantly renormalized by the atom-membrane coupling; (ii) the order parameters of the phases are calculated based on a Gaussian condensate profiles.

We plan to develop a variational technique to analyze the non-equilibrium quantum phase transition. The basic idea behind the variational method is to take a
trial function with a fixed shape, but with some free (time-dependent) parameters. Using a variational principle, we find a set of Newton-like second order ordinary differential equations for these parameters which characterize the solution. This technique has been used to study the non-equilibrium quantum phase transition of a hybrid atom-optomechanical system based 
on the one-component Gross-Pitaevskii equation coupled to a quantum oscillator. 

Lagrangian density of the hybrid system can be directly inferred from the effective Hamiltonian, reading
\begin{widetext}
\begin{eqnarray}
\mathcal{L} & =&\frac{1}{V}\left[\frac{i\hbar}{2}\left(\dot{\alpha}\alpha^{*}-\alpha\dot{\alpha}^{*}\right)-\hbar\Omega_{m}\alpha^{*}\alpha\right]+\frac{i\hbar}{2}\left(\Psi^\dagger\dot{\Psi}-\dot{\Psi}^\dagger{\Psi}\right)\nonumber\\
 & -&\left[\omega_{R}\left|\partial_x\Psi\right|^2+V\sin^{2}(x)\Psi^\dagger\Psi-2\sqrt{N}\lambda\alpha_1\sin\left(2x\right)\left|\psi_{1}\right|^{2}+\hbar\Omega\Psi^\dagger\sigma_x\Psi\right]-\frac{N}{4} \left[\left(g+g_{12}\right)\Psi^\dagger\Psi+\left(g-g_{12}\right)\Psi^\dagger\sigma_z\Psi\right]
\end{eqnarray}
\end{widetext}

Because the lattice potential can be approximately treated as a harmonic potential in each well,  we are motivated to write the order parameters of the model system as Gaussian profile
\begin{eqnarray}
\psi_{1}&=&\cos\theta\left[\frac{1}{\pi\sigma(t)^{2}}\right]^{1/4}e^{-\frac{(x-\zeta_{1}(t))}{2\sigma(t)^{2}}-i\kappa(t)x-i\beta(t)x^{2}}, \label{Try1}\\
\psi_{2}&=&\sin\theta\left[\frac{1}{\pi\sigma(t)^{2}}\right]^{1/4}e^{-\frac{(x-\zeta_{2}(t))}{2\sigma(t)^{2}}-i\kappa(t)x-i\beta(t)x^{2}}.\label{Try2}
\end{eqnarray}
\begin{figure*}
  \includegraphics[width=1.0\textwidth]{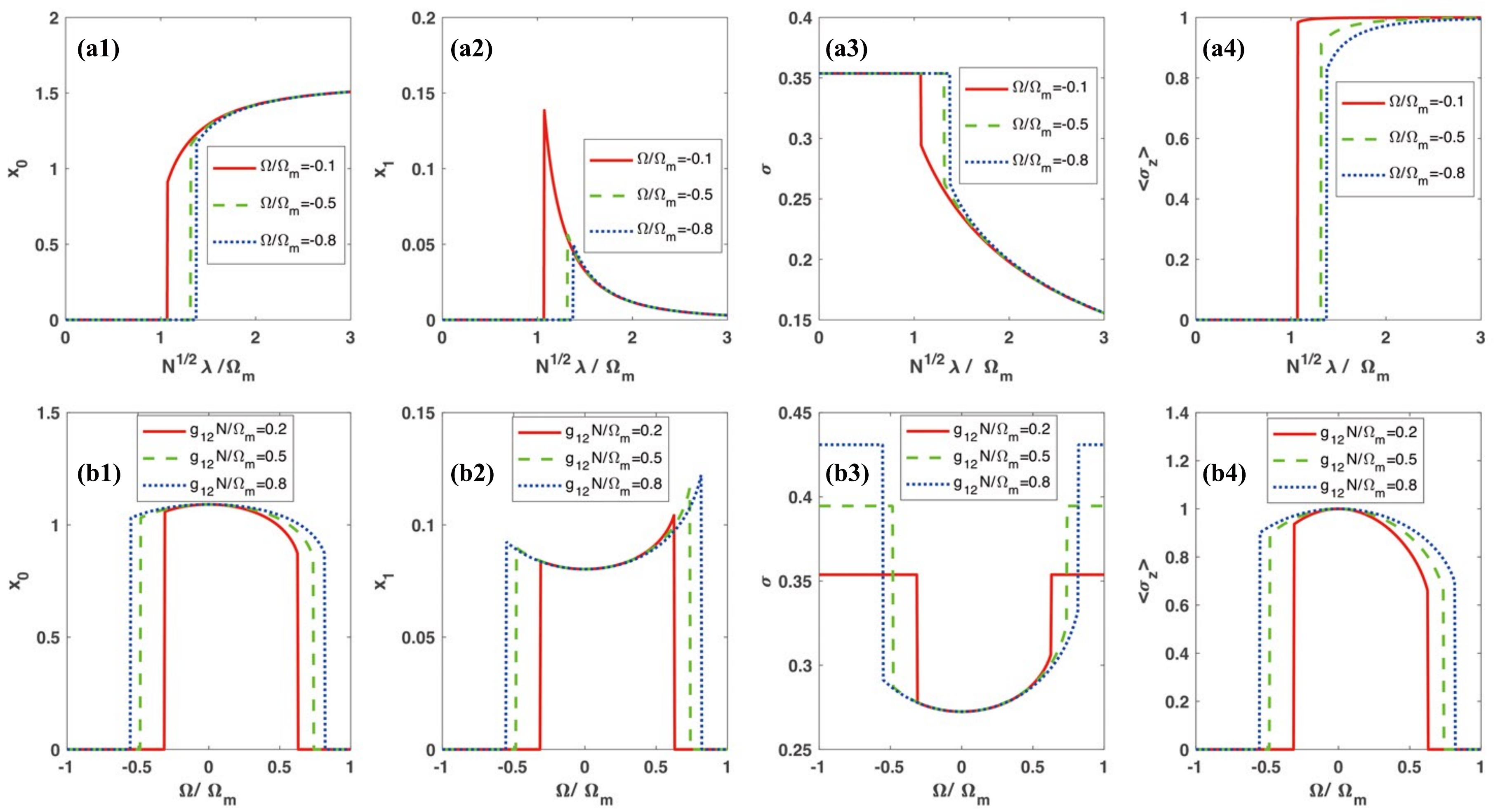}\\
\caption{Different parameters as a function of coupling strength (a1)-(a4) and Rabi frequency (b1)-(b4). Values of (a1),(b1) centered positions of two wavepackets, (a2),(b2) relative position, (a3),(b3) condensate width and (a4),(b4) $\left<\sigma_z\right>=n_1-n_2$ transform from zero to nonzero. Parameters  are used: $\Omega_m/\omega_R$=100, $gN/\Omega_m$=0.3, $V/\Omega_m$=2, $\gamma_R/\omega_R$=20, (a1)-(a4):  $g_{12}N/\Omega_m$=0.2 and (b1)-(b4) $\lambda N^{1/2}/\Omega_m$=1.2. }\label{f1}
\end{figure*}
In this work, we are limited into case : (i) two Gaussian wavepackets have the same width $\sigma(t)$ with the corresponding phase $\beta(t)$ and $\kappa(t)$; (ii) the centered positions of the two Gaussian wavepackets are different labelled by $\zeta_1(t)$ and $\zeta_2(t)$ respectively. 

With the help of the trial funcitons of Eqs. (\ref{Try1}) and (\ref{Try2}), we can proceed to obtain the Lagrangian of model system given by $L=\int \ensuremath{\mathcal{L}} dx$. Then, using Euler-Lagrange equation: $\frac{\partial}{\partial t}\frac{\partial L}{\partial\xi'}-\frac{\partial L}{\partial\xi}=0$
for different parameter $\xi$, we can arrive at the equations of motion for the different parameters $\beta$ and $\kappa$ as follows
\begin{eqnarray}
\dot{\zeta_{i}}&=&2\omega_R(\kappa+2\beta\zeta_{i}) (i=1,2),\\
\dot{\sigma}&=&4\omega_R\beta\sigma.
\end{eqnarray}
Next, we can proceed to obtain the equations of motion for  real number $\alpha_{1}$ and the imaginary number $\alpha_{2}$ respectively
\begin{eqnarray}
\Omega_{m}\alpha_{1}-\lambda\sqrt{N}\cos^{2}(\theta)e^{-\sigma^{2}}\sin(2\zeta_{1})+\alpha_{2}'&=&0 \label{La1} \\
\Omega_{m}\alpha_{2}-\alpha_{1}'&=&0\label{La2}
\end{eqnarray}
By substituting Eq. (\ref{La2}) to Eq. (\ref{La1}), we can obtain the equation of motion of $\alpha_1$ as follows
\begin{equation}
\frac{\alpha_{1}''+2\gamma\alpha_{1}'}{\Omega_{m}}=\lambda\sqrt{N}\cos^{2}\theta e^{-\sigma(t)^{2}}\sin(2\zeta_{1})-\Omega_{m}\alpha_{1}\label{aeq}
\end{equation} 

Inspired by Ref. \cite{Mann2018}, we calculate the effective energy functional of the atom part as follows
\begin{widetext}
\begin{eqnarray}
E&=&\tilde{\Omega}_{m}\alpha_{1}^{2}-2\lambda\sqrt{N}\cos^{2}(\theta)\alpha_{1}e^{-\sigma(t)^{2}}\sin(X_{0}+X_{1})-\frac{Ve^{-\sigma(t)^{2}}(\cos X_{0}\cos X_{1}-\cos(2\theta)\sin X_{0}\sin X_{1})}{2}\nonumber\\
&+&\frac{\omega_{R}}{2\sigma^{2}}+\Omega\sin(2\theta)e^{-\frac{X_1^{2}}{4\sigma(t)^{2}}}+V/2+\frac{Ng(\cos(4\theta)+3)+2Ng_{12}\sin^{2}(2\theta)e^{-\frac{X_1^{2}}{2\sigma^{2}}}}{8\sqrt{2\pi}\sigma}. \label{EnergyFunctional}
\end{eqnarray}
\end{widetext}
Note that two components BECs have different positions,  we can use centered position $X_{0}=\zeta_{1}+\zeta_{2}$ and relative position $X_{1}=\zeta_{1}-\zeta_{2}$. In the similar way of Ref. \cite{Mann2018}, the equations of motion related to the condensate can be written as
\begin{eqnarray}
\frac{X_{0}''}{2\omega_{R}}&=&-\partial_{X_0}E,\label{x0eq}\\
\frac{X_{1}''}{2\omega_{R}}&=&-\partial_{X_1}E,\label{x1eq}\\
\frac{\sigma''}{4\omega_R}&=&-\partial_{\sigma}E.\label{sigeq}
\end{eqnarray}

In determining the stationary-state phase diagram and the corresponding non-equilibrium phase transition of the energy functional (\ref{EnergyFunctional}), our strategy is based on the existence of four order parameters: the center-of-mass coordinate $X_0$, the relative  coordinate $X_1$, the width of the wave packet $\sigma$ and the longitudinal spin polarization $\langle\sigma_z\rangle$. Depending on the interplay among the three order parameters, we identify two phases in the stationary-state phase diagram as follows.

Phase I, localized symmetric phase, where both the center-of-mass coordinate $X_0$ and the relative  coordinate $X_1$ are equal to  zero and the longitudinal spin polarization $\langle\sigma_z\rangle=0$. The stationary state is the superposition of two same Gaussian functions centered in the lattice wells. 

Phase II, localized symmetry-broken phase, where both the center-of-mass coordinate $X_0$ and the relative  coordinate $X_1$ are equal to be nonzero and the longitudinal spin polarization $\langle\sigma_z\rangle\neq0$. The stationary state is the superposition of two same Gaussian functions with shifted atom configuration in the lattice wells. 

Below we drive the complete stationary-state phase diagram by numerically minimizing the energy functional (\ref{EnergyFunctional}). After the ansatz of Eqs. (\ref{Try1}) and (\ref{Try2}) are determined, we accordingly calculate the center-of-mass coordinate $X_0$, the relative  coordinate $X_1$, the width $\sigma$ and the longitudinal spin polarization $\langle\sigma_z\rangle\neq0$. In order to comprehensively reveal the effects of system's parameters, including the $\lambda$ and $\Omega$, on the non-equilibrium quantum phase transition from a localized symmetric state of atom cloud to a shifted symmetry-broken state, we have considered two cases for numerical analysis. (i) As is shown in Fig. (\ref{f1}) (a1)-(a4) two components hybrid system also has phase transition along with  increasing   coupling strength $\lambda\sqrt{N}$, centered position $X_0$ and polarized parameter $\left<\sigma_z\right>$ turn from zero to nonzero, besides relative position turn from zero to nonzero and then to zero. (ii) In Fig. (\ref{f1}) (b1)-(b4) when the coupling strength is fixed, Rabi frequency can  control phase transition, which brings a new method to cool the membrane.  Parameters have a jump at the critical point, which indicates 1st order transition, because two-component condensates are non-equal and this progress happens discontinuously.  If Rabi frequency is zero,  the 2nd component will vanish, for this reason  in Figs. 1(b1)-(b4) we ignore this case. For two components have a different role, when adjusting $\Omega$ the phase transition is asymmetric.

\section{Elementary excitation}

We now discuss how the stationary-state phase can be revealed in elementary excitations by solving Eqs. (\ref{aeq})-(\ref{sigeq}).
with the framework of the linear perturbation theory \cite{Nagy2008,Nagy2009,LiWu2015,ChenZhu2016}. After obtaining the stationary states of ($\alpha_{10}$,$X_{00}$,$X_{10}$,$\sigma_0$) in Eqs. (\ref{aeq})-(\ref{sigeq}), we proceed to calculate the collective spectrum by considering derivations from the stationary states in the form of $\alpha_{10}+\delta\alpha_{10}(t)$,$X_{00}+\delta X_{00}(t)$,$X_{10}+\delta X_{10}(t)$,$\sigma_0+\delta \sigma(t)$. Then we substitute the solutions to motion equations and rewrite differential equations in the form of a vector-matrix multiplications $\dot{v}=Mv$ with $v=(\delta\alpha_{10},\delta X_{00},\delta X_{10},\delta \sigma)$. With defining the following useful constants
\begin{eqnarray}
\omega_{1}&=&\lambda\sqrt{N}e^{-\sigma_{0}^{2}}\cos(X_{10}+X_{00}),\\
\omega_{2}&=&\lambda\sqrt{N}e^{-\sigma_{0}^{2}}\sin(X_{10}+X_{00}),\\
\omega_{a1}&=&e^{-\sigma_{0}^{2}}V\cos(X_{00})\cos(X_{10}),\\
\omega_{a2}&=&e^{-\sigma_{0}^{2}}V\cos(X_{00})\sin(X_{10}),\\
\omega_{b1}&=&e^{-\sigma_{0}^{2}}V\sin(X_{00})\sin(X_{10}),\\
\omega_{b2}&=&e^{-\sigma_{0}^{2}}V\sin(X_{00})\cos(X_{10}).
\end{eqnarray}
we finally obtain the matrix corresponding to the Bogoliubov-de Gennes \cite{LiWu2015,ChenZhu2016}, reading
\begin{widetext}
\begin{eqnarray}
M=\left(\begin{array}{cccccccc}
0 & 1 & 0 & 0 & 0 & 0 & 0 & 0\\
-\gamma^{2}-\Omega_{m}^{2} & -2\gamma & \omega_{1}\cos^{2}(\theta)\Omega_{m} & 0 & \omega_{1}\cos^{2}(\theta)\Omega_{m} & 0 & -2\sigma_{0}\omega_{2}\cos^{2}(\theta)\Omega_{m} & 0\\
0 & 0 & 0 & 1 & 0 & 0 & 0 & 0\\
8\omega_{1}\omega_{R} & 0 & -4\omega_{R}(2\alpha_{10}\omega_{2}+\omega_{a}) & 0 & \omega_{x01} & 0 & \omega_{x02} & 0\\
0 & 0 & 0 & 0 & 0 & 1 & 0 & 0\\
8\omega_{1}\omega_{R} & 0 & 4\omega_{R}(\omega_{b}-2\text{\ensuremath{\alpha_{10}}}\omega_{2}) & 0 & \omega_{x11} & 0 & \omega_{x12} & 0\\
0 & 0 & 0 & 0 & 0 & 0 & 0 & 1\\
-16\omega_{2}\sigma_{0}\cos^{2}(\theta)\omega_{R} & 0 & \omega_{\sigma1} & 0 & \omega_{\sigma2} & 0 & \omega_{\sigma3} & 0
\end{array}\right)\label{MatrixM}
\end{eqnarray}
\end{widetext}
\begin{figure}
  \includegraphics[width=0.8\textwidth]{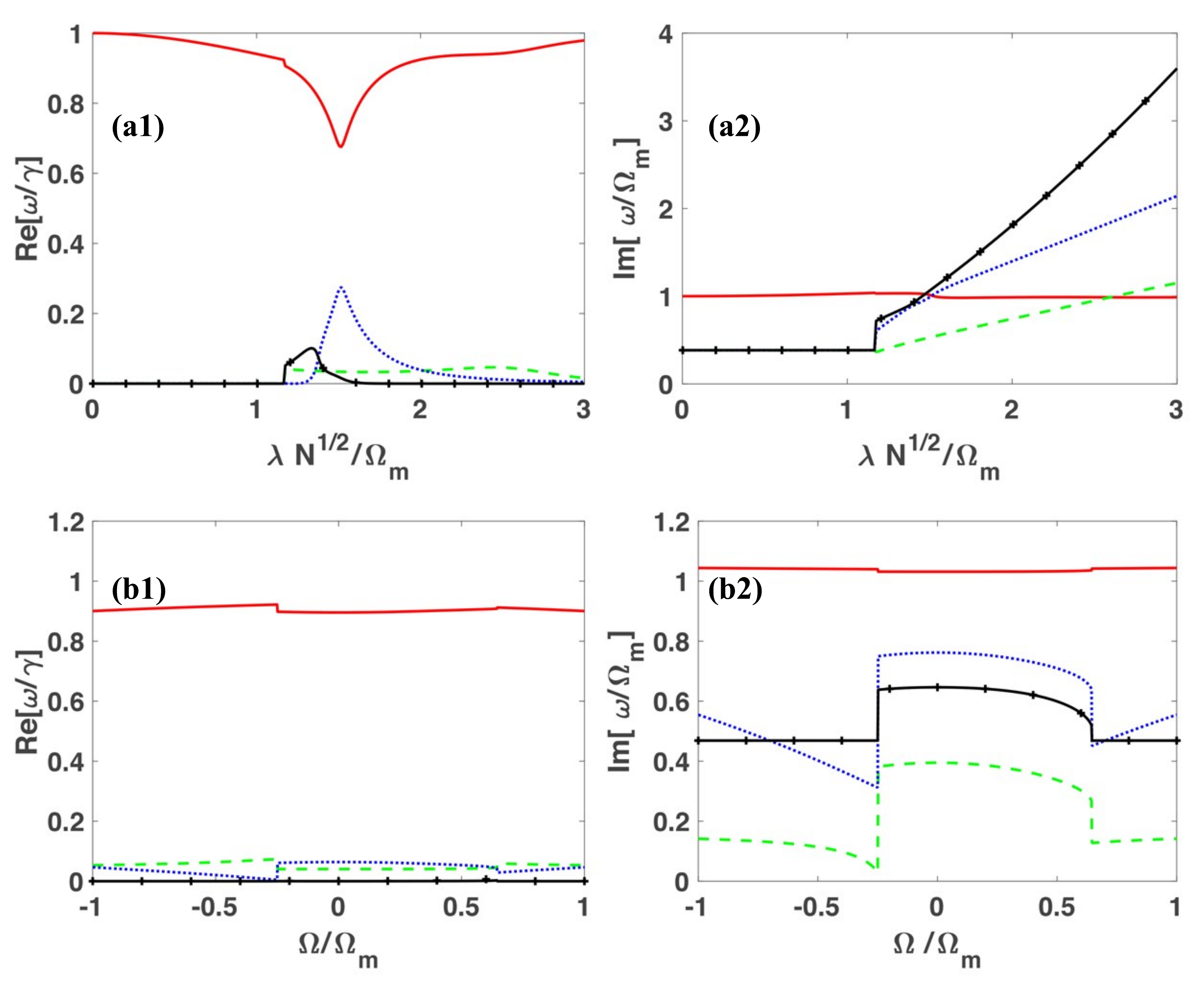}\\
\caption{Collective excitations of a spinor quantum gas in a lattice coupled to a membrane. (a1), (b1) Real and (a2), (b2) imaginary parts of excitations respectively. (a1)-(a2) show elementary excitation energy as function of coupling strength $\lambda$ and (b1)-(b2) show elementary excitation energy as function of Rabi frequency $\Omega$.  Parameters  are used: $\Omega_m/ \Omega_R$=100, $gN/\Omega_m$=0.3, $g_{12}N/\Omega_m$=0.2, $V/\Omega_m$=2, $\gamma/\omega_R$=20, (a1),(a2): $\Omega/\Omega_m$=-0.3 , and (b1),(b2): $\lambda \sqrt{N}/\Omega_m$=1.2 . 
}
\end{figure}
The real and imaginary parts of eigenvalues of the matrix (\ref{MatrixM}) define the eigenfrequencies and the decay rates respectively. In order to understand the effects of system's parameters, including the $\lambda$ and $\Omega$, on the nonequilibrium quantum phase transition in terms of the collective excitations, we consider the following two cases: (i) We first fix the values of $\Omega$ and check how the collective excitations change with varying the values of $\lambda$. As shown in Figs. 2 (a1) and (a2), the elementary excitations develop a jump at a critical point which is corresponding to the non-equilibrium quantum phase transition. (ii) As shown in Figs. 2 (b1) and (b2), similar jumps of the excitations occur when the $\Omega$ can induce  non-equilibrium quantum phase transition. As pointed out in Ref. \cite{Mann2018}, such kinds of jump in excitation can be used to probe the non-equilibrium quantum phase transition experimentally.

\section{Conclusion} \label{sec:conc}

Summarizing, motived by the experimental work \cite{Vochezer2018}, in which a novel kind of hybrid atom-optomechanical system has been realized by coupling atoms in a lattice to a membrane, we have further taken into account of the effects of the spinor degree of freedom of the atom part on the non-equilibrium phases of the hybrid atom-optomechanical system. In more details, a non-equilibrium quantum phase transition from a localized symmetric state of the atom cloud to a shifted symmetry-broken state, in particular, the effects of spinor degree of freedom on the non-equilibrium quantum phase transition are analyzed. The experimental realization of our scenario amounts to controlling two
parameters whose interplay underlies the physics of this work: the lattice strength $V$ and the effective atom-membrane coupling $\lambda$. 
With the state-of-the-art technology \cite{Vochezer2018}, the variation of $V$ and $\lambda$ can be reached by adjusting 
the laser power and cavity finesse. Moreover,  one can adjust the value of $\lambda$ independent on $V$ by applying a
second laser which is slightly misaligned with the first one generating an optical lattice of the same periodicity but shifted by $\pi/2$.

We remark that our theoretical framework of studying the non-equilibrium quantum phase transitions in this work is limited in the zero temperature. It's supposed that the backaction of the membrane vibration on the atoms may induce the possible temperature effect. In more details, the vibration of the membrane will lead to the shaking of the lattice by being mediated by the exchange of sideband photons of the lattice laser; as a result, the temperature of the atoms will increase. As estimated in our previous work \cite{Gao2019} with the typical experimental parameters,  the heating effect induced by the backaction of the membrane vibration on the atoms can be safely ignored by estimating the ratio between the energy scale of  the backaction of the membrane vibration on the atoms and chemical potential of the optically-trapped quantum gas as $\hbar \lambda /\mu\propto 10^{-2}$ \cite{Vogell2013,Vochezer2018}. We hope our work may induce the further experimental interests of quantum gases in a lattice coupled to a membrane with emphasis on the effects of the spinor degree of freedom. We emphasize here that the mean-field treatment of the hybrid atom-optomechanical system is limited to the Born-Markovian approximation of coupling between a membrane and the atoms at the zero temperature. For further investigations at the finite temperature or beyond Born-Markovian approximation, the path-integral Monte Carlo simulation should be a reliable theoretical framework.

\section{Acknowledgments}

We thank M. Reza Bakhtiari, Ying Hu, Chao Gao, Xianlong Gao, and Biao Wu for inspiring discussion. This work is supported by the NSFC of
China (Grant No. 11274315) and Youth Innovation Promotion
Association CAS (Grant No. 2013125).

\appendix
\section{The matrix elements in Eq. (\ref{MatrixM})}
The matrix elements in Eq. (\ref{MatrixM}) are given as follows:
\begin{widetext}
\begin{eqnarray}
\omega_{x01}/\omega_R&=&\frac{\sqrt{\frac{2}{\pi}}g_{12}N\cos(2\theta)e^{-\frac{X_{10}^{2}}{2\sigma_{0}^{2}}}(X_{10}^{2}-\sigma_{0}^{2})}{\sigma_{0}^{5}}+4\left(\omega_{b}-2\alpha_{10}\omega_{2}\right)+\frac{2\Omega\cot(2\theta)e^{-\frac{X_{10}^{2}}{4\sigma_{0}^{2}}}\left(X_{10}^{2}-2\sigma_{0}^{2}\right)}{\sigma_{0}^{4}}\\
\omega_{x02}/\omega_R&=&-\frac{\sqrt{\frac{2}{\pi}}g_{12}N\cos(2\theta)e^{-\frac{X_{10}^{2}}{2\sigma_{0}^{2}}}X_{10}\left(X_{10}^{2}-3\sigma_{0}^{2}\right)}{\sigma_{0}^{6}}+8\sigma_{0}\left(\omega_{b2}-2\alpha_{10}\omega_{1}\right)-\frac{2\Omega\cot(2\theta)e^{-\frac{X_{10}^{2}}{4\sigma_{0}^{2}}}X_{10}\left(X_{10}^{2}-4\sigma_{0}^{2}\right)}{\sigma_{0}^{5}}\\
\omega_{x11}/\omega_R&=&\frac{\sqrt{\frac{2}{\pi}}g_{12}Ne^{-\frac{X_{10}^{2}}{2\sigma_{0}^{2}}}\left(\sigma_{0}^{2}-X_{10}^{2}\right)}{\sigma_{0}^{5}}-4\left(2\alpha_{10}\omega_{2}+\omega_{a}\right)+\frac{\Omega\csc(\theta)\sec(\theta)e^{-\frac{X_{10}^{2}}{4\sigma_0^{2}}}\left(2\sigma_{0}^{2}-X_{10}^{2}\right)}{\sigma_{0}^{4}}\\
\omega_{x12}/\omega_{R}&=&\frac{\sqrt{\frac{2}{\pi}}g_{12}N\cos(2\theta)e^{-\frac{X_{10}^{2}}{2\sigma_{0}^{2}}}X_{10}\left(X_{10}^{2}-3\sigma_{0}^{2}\right)}{\sigma_{0}^{6}}+8\sigma_{0}\left(\omega_{a2}-2\alpha_{10}\omega_{1}\right)+\frac{\Omega\csc\theta\sec\theta e^{-\frac{X_{10}^{2}}{4\sigma_{0}^{2}}}X_{10}\left(X_{10}^{2}-4\sigma_{0}^{2}\right)}{\sigma_{0}^{5}}\\
\omega_{\sigma1}/\omega_{R}&=&4\sigma_{0}\left(-4\alpha_{10}\cos^{2}(\theta)\omega_{1}+\cos(2\theta)\omega_{a2}+\omega_{b2}\right)\\
\omega_{\sigma2}/\omega_{R}&=&\frac{4g_{12}N\sin^{2}(\theta)\cos^{2}(\theta)e^{-\frac{X_{10}^{2}}{2\sigma_{0}^{2}}}X_{10}\left(X_{10}^{2}-3\sigma_{0}^{2}\right)}{\sqrt{2\pi}\sigma_{0}^{6}}+4\sigma_{0}\left(-4\alpha_{10}\cos^{2}(\theta)\omega_{1}+\cos(2\theta)\omega_{b2}+\omega_{a2}\right)\nonumber\\
&+&\frac{\Omega\sin(2\theta)e^{-\frac{X_{10}^{2}}{4\sigma_{0}^{2}}}X_{10}\left(X_{10}^{2}-4\sigma_{0}^{2}\right)}{\sigma_{0}^{5}}\\
\omega_{\sigma3}/\omega_{R}&=&-\frac{\sqrt{\frac{2}{\pi}}gN(\cos(4\theta)+3)}{2\sigma_{0}^{3}}+\frac{g_{12}N(\cos(4\theta)-1)e^{-\frac{X_{10}^{2}}{2\sigma_{0}^{2}}}\left(2\sigma_{0}^{4}+X_{10}^{4}-5\sigma_{0}^{2}X_{10}^{2}\right)}{2\sqrt{2\pi}\sigma_{0}^{7}}\nonumber\\
&+&4\left(2\sigma_{0}^{2}-1\right)\left(4\alpha_{10}\cos^{2}(\theta)\omega_{2}-\cos(2\theta)\omega_{b}+\omega_{a}\right)-\frac{12\omega_{R}}{\sigma_{0}^{4}}-\frac{\Omega\sin(2\theta)e^{-\frac{X_{10}^{2}}{4\sigma_0^{2}}}X_{10}^{2}\left(X_{10}^{2}-6\sigma_{0}^{2}\right)}{\sigma_{0}^{6}}
\end{eqnarray}
\end{widetext}

\bibliography{AOM}
\end{document}